# Standards and Intellectual Property Rights in the Age of Global Communication - A Review of the International Standardization of Third-Generation Mobile System

Björn Hjelm

International Center for Standards Research
University of Colorado
Campus Box 530
Boulder, Colorado 80309-0530
bjorn.hjelm@ieee.org

*Abstract*-- When the European Telecommunications Standards Institute (ETSI) selected a radio access technology based on Wideband Code Division Multiple Access (WCDMA), sponsored by European telecommunications equipment manufactures Ericsson and Nokia, for its third-generation wireless communications system, a bitter dispute developed between ETSI and Qualcomm Inc. Qualcomm threatened to withhold its intellectual property on the CDMA technology unless the Europeans agreed to make the radio access technology backward compatible with the IS-95 standard, Qualcomm's favored version of CDMA. A dispute over key intellectual property rights of the CDMA technology also erupted between Ericsson and Qualcomm and both filed patent infringement in US Court. The dispute halted the development of air interface standards for third-generation mobile systems and troubled operators worldwide as well as the International Telecommunications Union (ITU).

I. INTRODUCTION

The vital importance of standards for compatibility and interoperability across a wide range of network industries is clear. Equally apparent is that the growing rewards from winning, or to some degree, controlling the outcome of the standardization process has made reaching consensus on standards much more difficult. Today, such agreements are essential if society is to enjoy the full benefits of global information and communications network, and of which wireless technologies will play a major role.[1]

A new mobile system for worldwide use is now being developed to enhance and supersede current second-generation digital systems. Referred to as third-generation mobile systems, it will provide universal personal communications to anyone worldwide. It will allow for broadband services such as high-speed data and wireless Internet access, full-motion video (videoconferencing), and a range of other multi-media functions and services.

The International Telecommunication Union (ITU) began its studies on global personal communications in 1985 resulting in a system referred to as International Mobile Telecommunications in the Year 2000 (IMT-2000), formerly known as Future Public Land Mobile Telecommunication System (FPLMTS). In parallel, the European Telecommunications Standards Institute (ETSI) developed a third-generation mobile system called Universal Mobile Telecommunications System (UMTS).

IMT-2000 is a family of systems that will let users roam worldwide with the same handset, and which will include UMTS as a subset. IMT-2000 will ensure that third-generation systems are globally compatible and provide uniform communications. The idea is to achieve this by encouraging all interested parties to work toward convergence of technologies that otherwise might compete against each other. However, that dream is somewhat clouded by a dispute between ETSI and leading European manufactures on one side and Qualcomm Inc. over the terms on which its intellectual property will be used in UMTS/IMT-2000 and the character of third-generation standards.

This paper explores the potential for conflict between the standards process and intellectual property rights (IPRs) in the New Global Economy by exploring the international standardization of third-generation mobile communications systems.

II. THE COMING OF AN INFORMATION AGE AND GLOBAL COMMUNICATION

With the globalization of many aspects of industries and the emerging of the Information Economy[2] (or "Digital

---

[1] *See* ITU Press Release, "ITU Works Towards Global Standards for Mobile Telecommunications," (October 1997) [hereinafter, "ITU/97-18"].

[2] *See* C. SHAPIRO AND H. R. VARIAN, INFORMATION RULES 1-3 (1999).

Economy"[3]), communication, and its different forms,[4] is now becoming truly global.[5] This means that the norm for communication is expanding beyond national boundaries and that distance is becoming less relevant.[6] A simple example of this phenomena is the growth of the Internet and have it is affecting the many aspects of the business and daily life. This also includes the aspect of managing intellectual property rights.[7]

Another aspect of this new age affecting intellectual property rights is the increasing rate of technological change and diffusion. Both the rate of technology changes and the speed at which new technologies become available and are used have increased substantially over the last 10 years. The shorter product life cycles resulting from this rapid diffusion of new technologies place a competitive premium on being able to quickly introduce new goods and services into the marketplace. Indeed, a key driver of the new economy is innovation and the theme is to "obsolete your own products"[8] as well as a shift from mass production to mass customization of goods and services. Innovation drives every aspects of economic and social life - a continual renewal of products, systems, processes, marketing and people.

The digitization of the new economy has paved the way for the information age. Today, information is in digital form that has changed how and at what speed information is transferred as well as improved the quality.[9]

Standards are serving as a foundation for enabling the global communication and this has also impacted both the development and economic value of standards.

### III. STANDARDS AND THE ECONOMIC ASPECTS OF STANDARDIZATION

Generically, a standard can be defined as a set of technical specification adhered to by a producer either tacitly or as a result of a formal agreement. In this paper, the definition will be restricted to the consideration of compatibility or interoperability standards. Though standards are used for many purposes the growing economic importance of standards has been revealed in high-profile cases such as Microsoft allege use of standards to dominate the world market for personal computers and, as described in this paper, also in other high technology industries such as the telecommunications industry.[10]

According to several literatures that have examined the economic processes affecting the formation of compatibility standards, a product emerges as a standard not only because of its inherent features but also because of the benefits deriving from a large installed base, termed network externalities. Direct network externalities are generated from the growing number of users adopting the product. Indirect network externalities are derived from the features not inherent to the product that increase with the number of adopters, and, consequently, add value to the core product. Compatibility can also offer significant cost savings through economies of scale.

Common standards can enhance levels of competition and international trade, but can also restrict competition and create trade barriers. Common standards can unify market requirements and foster competition amongst producers and service providers to benefit producers and consumers. Producers benefit from the enhanced value of their products ensuing from compatibility. Consumers benefit from being able to freely choose between different producers of compatible products. Common standards can lower trade barriers to international trade through global standards and promote innovation. On the other hand, divergent standards can create trade barriers and restrict competition by reducing variety. Standardization can hinder innovation through continued long-term acceptance of a standard that has been superseded by technically superior products or systems.

Once established, standards can dictate the nature of competition within product markets for many years to come. Control of the outcome can therefore yield

---

[3] *See* D. TAPSCOTT, DIGITAL ECONOMY 6 (1996) ("In the new economy, information in all its forms become digital – reduced to bits stored in computers and racing at the speed of light across networks.").

[4] As part of the diffusion, the dominant sector in the new economy is the new media, which are products of the convergence of the computing, communications, and content industries. The convergence is affecting all aspect of our daily life - the way we do business, work, play, live and probably even think. *See* TAPSCOTT, *supra* note 3, 59.

[5] A global economy is one in which goods, services, people, skills, and ideas move freely across geographic borders. To achieve strategic competitiveness in the global economy, a firm must view the world as its marketplace. Globalization is the spread of economic innovations around the world and the political and cultural adjustment that accompany this diffusion. Globalization encourages international integration and the range of opportunities for firms. *See* M. A. HITT, R. DUANE AND R. E. HOSKISSON, STRATEGIC MANAGEMENT – COMPETITIVENESS AND GLOBALIZATION 10 (1997).

[6] *See* F. CAIRNCROSS, THE DEATH OF DISTANCE (1997) at 1 ("The death of distance as a determinant of the cost of communicating will probably be the single most important force shaping society in the first half of the next century. Technological change has the power to revolutionize the way people live...It will alter, in ways that are only dimly imaginable...concepts of national borders and sovereignty, and patterns of international trade."). *See also* S. J. KOBRIN, *You Can't Declare Cyberspace National Territory* at 356 in BLUEPRINT TO THE DIGITAL ECONOMY (D. Tapscott ed., 1998) ("The emerged electronically networked global economy will affect how we are governed and how we live.").

[7] Global competition has increased performance standards in many dimensions, including those of quality of service, cost, productivity, product introduction time, and smooth, flowing operations. These standards are not static and require continuous improvement from a firm. Thus, competitive success will accrue only to those capable of meeting and exceeding global standards. *See* SHAPIRO AND VARIAN, *supra* note 2, 197 ("[T]he legal grant of exclusive rights to intellectual property rights…does not confer complete power to control information…[and there is] the issue of enforcement, a problem that has become even more important with the rise of digital technology and the Internet.").

[8] *See* TAPSCOTT, *supra* note 3, 5 ("If you're just developed a great product, your goal is to develop a better one that will make the first one obsolete. If you don't make it obsolete, someone else will").

[9] *See* TAPSCOTT, *supra* note 3, 48-49.

[10] *See* SHAPIRO AND VARIAN, *supra* note 2, 197 ("Strength…is measured along three primary dimensions: existing market position, technical capabilities, and control of intellectual property such as patents and copyrights.").

significant economic advantage on the sale of both core and related products.

IV. INTRODUCTION TO INTELLECTUAL PROPERTY RIGHTS

> *"Commerce…can seldom flourish long in any state which does not enjoy a regular administration of justice, in which people do not feel themselves secure in possession to their property, in which the faith of contracts is not supported by the law."*
>
> - Adam Smith, *Wealth of Nation*

Property law refers to a set of legal rules defining the rights of ownership of something having value, including the exclusive rights of property. There are two types of property, namely tangible and intangible. Tangible property relates to a physical object, while intangible property refers to a set of rights refers to the set of rights defined by law that are not related to physical objects.[11] Intellectual property is intangible and is created as well as protected based upon policy considerations "as to what types of intellectual activities should be encouraged."[12] In this paper, intellectual property is defined as the innovative or creative ideas of inventors, artists, or authors. Patents, copyright, trademark laws exist to provide incentives to create intellectual properties by ensuring that the owners of the intellectual properties maintain exclusive control over the ideas, at least for a certain period of time.

- Patents allow inventors the opportunity to recover their investment and the cost of creating and marketing inventions.
- Copyrights give authors control over the reproduction, dissemination, and public performance of their works.
- Trademarks assure consumers about product characteristics, such as quality.

The principle objective of intellectual property law is to encourage the development of new ideas and creations. As with all types of property, intellectual property may be sold, transferred, or otherwise disposed of.

*A. International Protection of Intellectual Property*

There is no international intellectual property law regime, in the sense that an innovator can obtain a universally recognized patent, copyright or trademark.[13] Instead, several international conventions ensure some degree of reciprocity among national systems. In addition, organizations such as the World Intellectual Property Organization (WIPO) collect information about, and promote harmonization of national laws.

Three multilateral treaties represent the primary sources of international private law protection for intellectual property. Those are the Berne Convention for the Protection of Literary and Artistic Works and the Universal Copyright Convention in the case of copyright, and the Paris Convention for the Protection of Industrial Property in the case of patent and trademark.[14]

In the United States, Section 301 0f the 1974 Trade Act, as amended by the 1988 Omnibus Trade Act, directs the United States Trade Representative to identify countries that fail to give adequate and effective intellectual property protections to the United States nationals.[15]

V. THE ROLE OF IPR IN THE STANDARDS PROCESS

Protection of IPRs is seen as necessary in creating sufficient incentives for companies to engage in innovation. At the same time, strong legal protection may exacerbate the problems of vested interest that complicate the formal standards process.[16]

As IPRs become more and more prominent in the global economy, the issues presented by the icorporation of patented technology into standards have become increasingly important.[17]

*A. The Changing Nature of Telecommunications Standardization*

In today's competitive telecommunications arena, it is not necessary to point out the huge strategic and economic significance of standards. Traditionally, telecommunications standards were established either on a national or international level.[18] The telecommunications sector has experienced major structural changes and eliminated much of the "natural" standards process on a national level[19] as well as created a more competitive international telecommunications field were standards serves as a mechanism to level the playing field for competition.

The growing number of players involved in telecommunication's standards has resulted in a more complex process of achieving standardization.[20] There is a growing concern within the industry that the current institutions responsible for creating standards are not able to deliver high-quality standards at the pace demanded by the market. The potential negative effect of strong IPRs or abuse of the process[21] has also been the subject of much concern in the formal standards arena and the introduction of digital technology has heightened the importance of IPRs. Standards-setting procedures play an important role in influencing these effects.[22]

---

[11] *See* D. A. GREGORY, C. W. SABER, AND J. D. GROSSMAN, INTRODUCTION TO INTELLECTUAL PROPERTY LAW 1 (1994).
[12] *Id.* at 3.
[13] *See* P. B. STEPHAN III, D. WALLACE, JR., AND J. A. ROBIN, INTERNATIONAL BUSINESS AND ECONOMICS 397-461 (1993).
[14] *See* P. GOLDSEIN, COPYRIGHT, PATENT, TRADEMARK AND RELATED STATE DOCTRINES Chapter 5 (1993).
[15] *See* D. B. Newman, Jr., *Intellectual Property Reforms and International Trade*, 27 IEEE COMMUNICATIONS MAGAZINE 41-42 (January 1989).
[16] *See* SHAPIRO AND VARIAN, *supra* note 3, at 16 ("Very often, support for a new technology can be assembled in the context of a formal standard-setting effort…both Motorola and Qualcomm have sought to gain competitive advantage, not to mention royalty income, by having their patented technologies incorporated into formal standards for modems and cellular telephones.").
[17] *See* K. Krechmer, *Communications Standards and Patent Rights: Conflict or Coordination?*, TIA STAR (1997).
[18] The dominant telecommunications carrier set standards on a de facto basis and coordination problems were eliminated because of the carrier's monopoly position. Compatibility between autonomous national networks was achieved through bilateral government agreements.
[19] Market liberalization and rapid technological change have eliminated the dominant position previously experienced by local monopoly telecommunication carriers.

*B. Standards and IPR - A Fundamental Dilemma*

The dilemma for policy makers is how far IPRs should be overridden in the public interest of common standard.

Though standardization and IPRs share the same broad economic objective to ensure that society benefits the fullest from innovation, their approaches differ. IPRs are oriented toward producers and reflect the trade-off between the need to create sufficient incentives for innovation and the public good nature of an innovation once it has been discovered. Standardization is consumer oriented and seeks to encourage a common platform whereby users benefit from enhanced competition and trade.

A small relief to the dilemma can be found in the IPR licensing provision. A license is a permission to carry out acts otherwise prohibited by virtue of the exclusive rights and IPR owner has. Licensing is a power solely reserved to the IPR holder and can be described as:

- An IPR owner is not obligated to license out an IPR
- An IPR license may discriminate amongst potential licensees.
- Where an IPR holder does issue a license, it is entitled to secure any such monetary or other considerations[23] that it is able to extract from the licensee(s) selected.

In standardization, the IPR owners will typically offer up licensees (1) for a fair sum, (2) on a reasonable terms and conditions, and (3) on a non-discriminatory basis.

*C. ITU IPR Policy*

Generally, the IPR policies adopted by standards developing organizations aim to how disclose relevant intellectual property rights during the initial standards developong process as well subsequesnt licensing efforts.

From Resolution ITU-R 1-2, the "Statement on Radiocommunication Sector patent Policy" covering "code of practice" regarding intellectual property rights (patents) covering, in varying degrees, the subject matters of ITU-R Recommendations[24] can be summarized as:

§ 1   "The ITU is not in a position to give authoritative or comprehensive information about evidence, validity or scope of patents or similar rights, but it is desirable that the fullest available information should be disclosed. Therefore, any Radiocommunication Sector Member organization putting forward a proposal for recommendation should, from the outset, draw the attention of the Director of the Radiocommunication Bureau to any known patent or to any known pending patent application, either their own or other organizations, although the Director of the Radiocommunication Bureau is unable to verify the validity of any such information."

§ 2   "If an ITU-R Recommendation is developed and such information as referred to in § 1 has been disclosed, three different situations may arise:"

§ 2.1   "The patent holder waives his rights; hence, the Recommendation is freely accessible to everybody, subject to no particular conditions, no royalties are due, etc."

§ 2.2   "The patent holder is not prepared to waive his rights but would be willing to negotiate licenses with other parties on a non-discriminatory basis on reasonable terms and conditions. Such negotiations are left to the parties concerned and are performed outside the ITU-R."

§ 2.3   "The patent holder is not willing to comply with the provisions of either § 2.1 or § 2.2; in such case, no Recommendation can be established."

§ 3   "Whatever case applies (§§ 2.1, 2.2 or 2.3), the patent holder has to provide a written statement to be filed at the Radiocommunication Bureau. This statement must not include additional provisions,

---

[20] National and international standards bodies now face a coordination problem posed by an ever growing membership as well as an increased breadth and complexity of work undertaken due to rapid technological advances.
[21] *See* J. Kipnis, *Beating the System: Abuses of the Standards Adoption Process*, 38 IEEE COMMUNICATIONS MAGAZINE 102 (July 2000) ("Companies can word their patent statements to appear to comply with the bylaws of organization while not actually doing so; interpret their statements to provide a basis to refuse to odder licenses to certain parties; and hide their intellectual property until after a standard has been adopted and has gained widespread acceptance").
[22] *See* R. P. Feldman and M. L. Rees, *The Effect of Industry Standard Setting on Patent Licensing and Enforcement*, 38 IEEE COMMUNICATIONS MAGAZINE 116 (July 2000) ("Industry standard-setting proceedings and the rules implemented to guide those proceedings play an important part in the licensing and enforcement of patents in many technological fields.").
[23] In March 30, 1998, Qualcomm, Inc. and Philips Consumer Communications LP (PCC) entered into a CDMA royalty-bearing, cross-license agreement together with Philips Electronic NV. Under the terms of the agreement, Qualcomm granted PCC a license under certain of its patents to develop, manufacture and sell CDMA and WCDMA base units, while Philips granted Qualcomm a license under certain of Philip's patents for its own use and sale of CDMA products.

[24] ITU-R Recommendations are non-binding international documents. Their objective is to ensure the rational, equitable, efficient and economical use of radio-frequency spectrum and satellite orbits or to recommend on various radiocommunication matters. To meet this objective, which is in the common interests of all those participating in radiocommunications it must be ensured that Recommendations, their applications, use, etc. are accessible to everybody. It follows therefore that a commercial (monopolistic) abuse by a holder of a patent embodied fully or partly in a Recommendation must be excluded. To meet this requirement in general is the sole objective of the code of practice. The detailed arrangements arising from patents (licensing, royalties, etc.) are being left to the parties concerned, as these arrangements might differ from case to case.

conditions, or any other exclusion clauses in excess of what is provided for each case in §§ 2.1, 2.2 and 2.3."

## VI. THE DEVELOPMENT OF THIRD-GENERATION MOBILE SYSTEM

A new mobile system for worldwide use is now being developed to enhance and supersede current second-generation mobile systems.[25] Third-generation mobile systems are driven by the vision of *information at any time, at any place, in any form* and aim at providing universal personal communications to anyone worldwide. Research efforts have been aligned with efforts in the International Telecommunication Union (ITU) and other bodies to find standards and recommendations which ensure that mobile communications of the future have access to multimedia capabilities and service quality similar to the fixed network.[26]

### A. International Telecommunications Union

At the end of 1985, the International Radiocommunications Consultative Committee (CCIR) established a group, now ITU Radiocommunications Sector (ITU-R) Task Group 8/1 (ITU-R TG 8/1),[27] to identify the needs of Future Public Land Mobile Telecommunication System (FPLMTS),[28] later referred to as International Mobile Telecommunications in the Year 2000 (IMT-2000),[29] The ITU World Administrative Radio Conference in 1992 (WARC-92)[30] identified frequency spectrum on a worldwide basis for the satellite and terrestrial components of IMT-2000 around 2 GHz,[31] later revised at WRC-95 to include mobile satellite service (MSS) which will provide the satellite component of third-generation mobile system.[32]

WARC-92 also adopted Resolution 212, providing the general framework for IMT-2000 standards development and system implementation. The standardization of third-generation mobile system within ITU involved both the Radiocommunication Sector (ITU-R) and Telecommunication Standardization Sector (ITU-T).[33] The approach was designed to capitalize as much as possible on common radio-related functions in the many different radio operating environments.

### B. European Research and Standardization

In 1990, the European Telecommunications Standards Institute (ETSI) established an "ad hoc" group on UMTS, later Subgroup 5 of the Special Mobile Group (SMG5), which focused on the critical points to be studied for systems suitable for providing personal communication services to people on the move.[34] At the end of 1995, the UMTS work program and responsibilities were reconstructed due to the further influence of the market and the general ETSI restructuring of its technical bodies. The SMG technical committee was given overall responsibility for UMTS standardization.[35]

Since the end of 1998, ETSI's standardization of third-generation mobile system has been carried out in the 3rd Generation Partnership Project (3GPP).[36]

### C. Third-Generation Partnership Projects

The concept initiated by ETSI in the beginning of 1998, the 3GPP is a consortia of the five global standards development organizations (SDOs) ETSI, Association of Radio Industries and Businesses (ARIB), Committee T1 (T1), Telecommunications Technology Association (TTA) and Telecommunications Technology Committee (TTC). The 3GPP have agreed to co-operate for the production of Technical Specifications for a Third-Generation Mobile System based on the evolved GSM core networks and the radio access technologies supported by the five SDOs.

---

[25] Second-generation mobile systems include the time division multiple access (TDMA) systems Global System for Mobile communications (GSM) and IS-136 standard as well as the code division multiple access (CDMA) system IS-95 standard.
[26] *See* ITU/97-18, *supra* note 1.
[27] The main goal of the group was to define the services that could potentially be delivered by radio, with particular reference to access and terminal mobility aspects. This work had a close relationship with ITU-Telecommunications Sector (ITU-T) studies on Universal Personal Telecommunications (UPT) which is a service concept aimed at the provision of full personal mobility. *See* K. Asatani, *Standardization of Network Technologies and Services*, 32 IEEE COMMUNICATIONS MAGAZINE 86-91 (July 1994).
[28] *See* M. H. Callendar, *Future Public Land Mobile Telecommunication Systems*, 1 IEEE PERSONAL COMMUNICATIONS 18-22 (Fourth Quarter 1994).
[29] *See* Special issue on IMT-2000: Standards Efforts of the ITU, 4 IEEE PERSONAL COMMUNICATIONS (August 1997).
[30] The ITU, through the Radio Regulations (RR), which are the basis of a treaty representing international law to be observed by the signatory states, has been responsible for global harmonization of radio spectrum resources. The ITU RR are reviewed and updated as necessary at the World Radio Communication Conferences (WRCs).
[31] This regulatory provision is part of the ITU RR footnote S5.388, which establishes that "the bands 1885-2025 MHz and 2110-2200 MHz are intended for use on a worldwide basis by administrations wishing to implement FLPMTS."
[32] At present, spectrum requirements for terrestrial component of the system are estimated to be around 500 MHz, with additional spectrum likely to be identified in the bands below 3 GHz. The issue of spectrum for third-generation mobile system will be considered again at WRC-2000.
[33] ITU-R Study Group 8/1 - IMT-2000; ITU-T Study Group 2 - Services & Operations, Study Group 3 - Charging &Tariffs, Study Group 4 – Management, Study Group 7 – Data (including Security), Study Group 11 – Signaling & Protocols, Study Group 12 – Performance, Study Group 13 – Networks, and Study Group 16 – Multimedia Services & Speech Coding.
[34] E. Damosso and G. De Brito, *COST 231 Achievements as a Support to the Development of UMTS: A Look into the Future*, 34 IEEE COMMUNICATIONS MAGAZINE 90-96 (February 1996).
[35] Subtechnical committee SMG1 is responsible for UMTS service aspects, SMG2 is responsible for the specification of UMTS generic radio access, SMG3 for GSM Core Network evolution, and SMG12 for overall UMTS architecture.
[36] *See* P. Chaudhury, W. Mohr, and S. Ono, *The 3GPP Proposal for IMT-2000*, 37 IEEE COMMUNICATIONS MAGAZINE 72-81 (December 1999).

The 3rd Generation Partnership Project 2 (3GPP2) is an effort spearheaded by the International Committee of the American National Standards Institute's (ANSI) board of directors to establish a 3G Partnership Project for evolved ANSI-41 networks and related radio transmission technologies (RTTs).

*D. The International Standardization of the Air Interface for Third-Generation Mobile System*

Since the work started in the standardization bodies ITU and ETSI, third-generation activities have formed an umbrella for advanced radio system developments. The IMT-2000 radio interface specifications is being developed jointly by various manufactures, operators, organizations, and standardization bodies that participate in the work of the ITU.[37] A formal request by the ITU-R for submission of candidate radio transmission technologies (RTTs) for IMT-2000 was distributed by the ITU, with a closing date of June 1998.[38] The evaluation of these proposals is based on Rec. ITU-R M.1225[39] and was scheduled to be completed end of March 1999.[40]

The main objectives for the IMT-2000/UMTS air interface were:

- Full coverage and mobility for 144 Kb/s, preferably 384 Kb/s.
- Limited coverage and mobility for 2 Mb/s.
- High spectrum efficiency compared to existing systems.
- High flexibility to introduce new service.

In Europe, significant progress has been made since late 1980s towards the development of future generations of mobile communication concepts, systems and networks through a number of European Union funded R&D projects.[41]

On January 29, 1998, ETSI selected the basic technology for the UMTS terrestrial radio access (UTRA) system. This decision contained the following key elements:

- For the paired bands 1920 – 1980 and 2110 – 2170 MHz wideband code-division multiple access (WCDMA) shall be used in frequency-division duplex (FDD) operation.
- For the unpaired bands of total 35 MHz time-division code-division multiple access (TD-CDMA) shall be used in time-division duplex (TDD) operation
- Parameters shall be chosen to facilitate easy implementation of FDD/TDD dual-mode terminals.

In March 1998, the TIA (Telecommunications Industry Association) TR45.5 committee, responsible for IS-95 standardization, adopted a framework for wideband CDMA (i.e. CDMA2000) backward compatible to IS-95 systems. The main difference between UTRA and CDMA2000 systems, supported by Qualcomm, were chip rate, downlink channel structure, and network synchronization.[42] In general, the difference is based on considerations of backward compatibility to second-generation systems.[43] The majority of the proposals submitted to ITU-R for candidate RTTs on the IMT-2000 terrestrial component used CDMA as the choice for multiple access technique.[44]

VII. IPR AND THE STANDARDIZATION OF THIRD-GENERATION MOBILE SYSTEMS

The IPR issue concerning the right to key CDMA technology patents made the standardization of a international third-generation mobile system very complex and resulted in delayed decision regarding key standards in IMT-2000[45] as well as allegation of international trade violations.

In late April 1998, Qualcomm informed ETSI that unless the UMTS proposal provided backward compatibility to the IS-95 standard, it would deny access to the intellectual property it claimed was essential to wideband CDMA development.

*A. Ericsson v. Qualcomm*

Since 1995, Ericsson, Inc. and Qualcomm, Inc. had been involved in litigation over CDMA IPRs and technological patents. It was Qualcomm's contention that the proposed wideband CDMA standard (WCDMA) by ETSI was specifically designed to exclude Qualcomm technology from the proposed standard.[46] In doing so, Qualcomm

---

[37] *See* R. D. Carsello, R. Meidan, S. Allpress, F. O'Brian, J. A. Tarallo, N. Ziesse, A. Arunachalam, J. M. Costa, E. Berruto, R. C. Kirby, A. Maclatchy, F. Watanabe, and H. Xia, *IMT-2000 Standards: Radio Aspects*, 4 IEEE PERSONAL COMMUNICATIONS 30-40 (August 1997).

[38] *See* ITU-R Rec. M.1225, *Guidelines for Evaluation of Radio Transmission Technologies (RTTs) for IMT-2000* (1997).

[39] *See* ITU-R, *Request for Submission of Candidate Radio Transmission Technologies (RTTs) for IMT-2000/FPLMTS Radio Interface*, CIRC. LETT. 8/LCCE/47 (April 4, 1997).

[40] *See IMT-2000 The Global Standards for Personal Communications*, ITU PUBLICATIONS 15 (1998).

[41] The European Commission has sponsored research programs such as Research and Development of Advanced Communication Technologies in Europe (RACE I and RACE II) and Advanced Communications Technology and Services (ACTS) in order to stimulate research on future mobile communication. In particular, the projects CODIT (evaluation of code-division multiple access, CDMA) and ATDMA (evaluation of time-division multiple access, TDMA) within RACE II as well as FRAMES (evaluation of future radio wideband multiple access systems) within ACTS have been very important for the development of the terrestrial radio access system for UMTS. *See also* J. S. Dasilva, B. Arroyo-Fernàndez, B. Barani, D. Ikonomou, *European Third- Generation Mobile Systems*, 34 IEEE COMMUNICATIONS MAGAZINE 68-83 (October 1996).

[42] *See* T. Ojanperä and R. Prasad, *An Overview of Air Interface Multiple Access for IMT-2000/UMTS*, 36 IEEE COMMUNICATIONS MAGAZINE 82-95 (September 1998).

[43] UTRA is backward compatible to the European GSM and Japanese PDC system, while CDMA2000 is backward compatible to the IS-95 standard.

[44] By the end of June 1998, there were 10 proposals submitted to ITU-R for candidate RTTs on the IMT-2000 terrestrial component for from the United States, Europe, Japan, China, and Korea.

[45] ITU indicated in the beginning of December 1998 that they would only be able to consider RTT technologies for IMT-2000 that were based on TDMA technology if the dispute surrounding IPR of CDMA proposals was not resolved before the end of 1998.

[46] UTRA is designed to be compatible with GSM, the second-generation mobile system widely deployed in Europe and other parts of the world, and

charged that this constituted a violation of free trade laws between the U.S. and Europe. Qualcomm claimed patent infringement on key technologies needed for WCDMA. In October 1998, the company formally invoked claims of intellectual property rights to five RTT proposals pending before the ITU and stated that it would refuse to license any ITU terms unless a converged and IS-95 compatible standard resulted from the IMT-2000 initiative. Qualcomm expected a harmonized standard to greatly increase its IPR licensing revenue,[47] an important source of profit,[48] and that this would open the European market.

It was Ericsson's position that the current WCDMA standard did not infringe on any of the IPRs claimed by Qualcomm. In a lawsuit filed by Ericsson that was scheduled for early 1999, Ericsson claimed Qualcomm had infringed on several of Ericsson's CDMA patents. Although no other companies own CDMA patents, Ericsson claimed they owned several of the key patents of the IS-95 standard involving soft handoff and macrodiversity. Ericsson claimed that Qualcomm refusal to license CDMA IPRs was simply a way to protect embedded investment and expand market share. Ericsson had announced that it would move forward on WCDMA development without licenses from Qualcomm and not license its WCDMA and CDMA2000 patents to companies that did not reciprocate.[49]

### B. International Trade Implications

Different countries provide different levels of intellectual property protection, and this can have significant effect on international trade.[50] Governments often regulate the production and distribution of products and goods. Sometimes such standards can limit trade.

The dispute between Qualcomm and ETSI turned into a contentious trade issue between the U.S. and European Union (EU). On June 4, 1998, John Major, Executive Vice President at Qualcomm, told the U.S. House Subcommittee on Technology that the technology being adopted in Europe would not provide an evolutionary path for current the IS-95 standard, effectively forcing operators to deploy entirely new third-generation system rather than leveraging existing investment. In September 1998, the U.S. Senate passed a resolution calling for a harmonized global third-generation standard, compatible with all legacy wireless system. Later the same year, the Clinton administration stated it was prepared to vigorously engage EU over trade barriers to global competition in the third-generation mobile phone market. The U.S. claimed that since Europe has chosen a *de facto* WCDMA standard for third-generation mobile systems, Europe was effectively keeping U.S. based technologies from gaining market entry to the EU.[51] European officials responded to these charges by stating that they did not intend to mandate a single third-generation wireless standard prior to the conclusion of the ITU standards process.[52] In an EU decision on UMTS it did neither define any technological content nor did it establish UMTS as an exclusive standard.[53]

### C. The Effects on Developing Global Third-Generation Mobile System Standards

The standardization of the air interface for third-generation mobile system was delayed and operators around the world decided to join forces in supporting one standard, compatible to all second-generation mobile systems.[54]

During the ITU-R Task Group 8/1 meeting in March 1999, it was agreed on the way to proceed with the standardization of IMT-2000. The decisions reached at Fortaleza, Brazil, provide essentially a single flexible standard with a choice of multiple access methods which include code-division multiple access (CDMA), time-division multiple access (TDMA) and combined TDMA/CDMA, all potentially in combination with space-division multiple access SDMA, to meet the many different mobile operational environments around the world.[55]

Participants agreed that the further development of the more detailed IMT-2000 radio interface recommendations within the ITU should continue to aim at minimizing the impact of flexibility within the IMT-2000 standard on users through maximizing commonality and ease of digital implementation in a hand-held mobile unit. It was also agreed that IMT-2000 radio interfaces should include the

---

not IS-95, the CDMA-based second-generation mobile system pioneered by Qualcomm.
[47] In 1998, the CDMA license royalties comprised 4.1% of the Qualcomm's sales.
[48] *See* C. Carlson, *An Introduction to 3G: What's the Ballyhoo?*, WIRELESS WEEK (October 27, 1998) <http://www.wirelessweek.com/3G/intro.htm>.
[49] Ericsson is said to be fully prepared to grant licenses to patents on fair, reasonable, and non-discriminatory terms subject to conditions of reciprocity which are required to create fairness in a multi-standard environment.
[50] *See* S. Husted and M. Melvin, INTERNATIONAL ECONOMICS Chapter 7 (1997).
[51] *See* Testimony by Kevin Kelly before the Sub-Committee on Trade of House Committee Ways and Means (July 29, 1998) <http://www.house.gov/ways_means/trade/testimony/7-28-98/7-28kell.htm>.
[52] In response to a December 19 letter signed by US Secretary of State Madeleine Albright, US Trade Representative Charlene Barshefsky, US Secretary of Commerce William Daley and Federal Communication Commissioner William Kennard, EU Telecommunications Commissioner Martin Bangemann firmly rejected US allegations that the EU is raising barriers to US firms in the third-generation mobile communications market. EU policy, answered Commissioner Bangeman, is to have market demand met by a broad competitive offering of mobile multimedia services, fully in line with the EU regulatory framework and its WTO obligations. Bangemena underlined that the EU fully supports global harmonization of third-generation technology standards via the ITU, but that the decisions on this must be an industry-led process of technical and commercial considerations.
[53] *See* COMMON POSITION (EC) /98 *Adopted by the Council Concerning the Coordinated Introduction of a Third -Generation Mobile and Wireless Communications System (UMTS) in the Community* (September 24, 1998).
[54] *See* NTT DoCoMo International Press Release, *Standardization of Third-Generation Mobile Communications System* (December 18, 1998) <http://www.nttdocomo.com/pr/pt981218.htm>.
[55] Although second-generation mobile systems involve both TDMA and CDMA technologies, very little use is currently being made of SDMA. The advent of adaptive antenna technology linked to systems which have been designed to optimize performance in the space dimension should significantly enhance the performance of future systems.

capability of operating with both of the major third-generation core networks currently under development. The key characteristics by themselves did not constitute an implementable specification but established the major features and design parameters that would make it possible to develop the detailed specs (to be referred to as IMT.RSPC and later ITU-R Rec. M.1457) between June and November 1999.

The flexible approach represented the only option on which consensus could be achieved and work could proceed. The meeting nonetheless agreed to strongly encourage the various operators for in their efforts to achieve a minimum set of radio interfaces, covering operators needs having the least possible impact on mobile terminals so that the user is unaware of the technology which provides the services chosen, and thus meet the widely endorsed IMT-2000 objectives. The operators were also requested to provide comments on the flexibility provided in the key characteristics as approved in Fortaleza and possibly to add further information on operational scenarios those operators' face around the world. Operators were also urged to convey their views to the radio transmission technology proponents to facilitate the rapid development of ITU-R Recommendations for IMT-2000.

## VIII. CONCLUSIONS

There is clearly a potential conflict between standards and IPRs. The dilemma is that while strong legal protection of IPRs can intensify the difficulties of reaching standards agreement, especially on an international level. The extent to which it is possible to preserve IPRs and promote standardization is a dilemma constantly faced by the formal standards bodies. The use of existing legal framework offers only a partial and increasingly inadequate solution. As the value of standards grows, so are the incentives for companies to use IPRs as a strategic tool with which to attempt to control the pace and direction of the standards process. The analysis of this paper has highlighted the international importance of IPRs and that it threatens to slow, or in some case halt, the standards process.

## IX. PROLOGUE

On March 25, 1999, Ericsson and Qualcomm announced that they had entered into a series of definitive agreements that resolve all disputes globally between the companies relating to code-division multiple access (CDMA) technology. Under the agreements, Ericsson and Qualcomm agree to jointly support a single world CDMA standard with three optional modes for the third-generation of wireless communications, enter into cross licenses for their respective patent portfolios and settle the existing litigation between the companies. The cross licenses are royalty bearing for CDMA subscriber units sold by either party. In addition, Ericsson purchased Qualcomm's terrestrial CDMA wireless infrastructure business, including its R&D facilities, located in San Diego, Calif. and Boulder, CO, and assumed selected customer commitments, including a portion of vendor financing obligations, related assets and personnel. The agreement settled the litigation between Ericsson and Qualcomm and provided cross licensing of IPRs for all CDMA technologies, including the IS-95 standard, WCDMA (i.e. ETSI's UTRA) and CDMA2000. Qualcomm also received rights to sublicense certain Ericsson patents, including the patents asserted in the litigation, to Qualcomm's Application Specific Integrated Circuits (ASICs) customers.

At the 17th meeting of the International Telecommunication Union group of radio experts on IMT-2000 (ITU-R Task Group 8/1) which met in Beijing in June 1999, "Qualcomm and Ericsson both submitted formal statements concerning the resolution of the Intellectual Property Rights problems on CDMA2000 and W-CDMA technologies which indicate that all disputes are globally resolved between the two companies. The statements also confirm the companies' commitment to license their essential patents for a single CDMA standard or any of its modes on a fair and reasonable basis, free from unfair discrimination."[56]

At the ITU-R Task Group 8/1 meeting on November 5, 1999, radio interface specifications for IMT-2000 incorporating the flexibility required were finally approved (Figure 1).[57]

FIGURE 1. IMT-2000 RADIO INTERFACE SPECIFICATIONS.

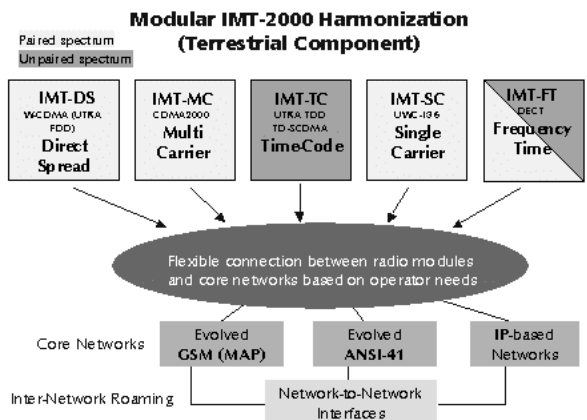

---

[56] See ITU Press Release, *Major progress in Beijing on standardization of IMT-2000* (June 15, 1999).
[57] See ITU Press Release, *ITU gears up to deliver future proof solutions to support seamless global roaming across networks: six network-related standards agreed today* (December 10, 1999).